\documentstyle[preprint,revtex]{aps}

\begin{document}
\draft
\preprint{ UBCTP 92-022  \\June 1992}
\begin{title}
{\bf Induced QCD and Hidden Local Z$_{\bf N}$ Symmetry}
\end{title}

\vskip 1in

\author{Ian I. Kogan\footnote{Permanent address, ITEP, Moscow, Russia},
Gordon W. Semenoff and Nathan Weiss}

\vskip 15pt

\begin{instit}
Department of Physics, University of British Columbia,\\
Vancouver, British Columba, Canada  V6T 1Z1
\end{instit}

\vskip .5in

\begin{abstract}
\vskip .5in
We show that a lattice model for induced lattice QCD which was
recently proposed by Kazakov and Migdal has a $Z_N$ gauge symmetry
which, in the strong coupling phase, results in a local confinement
where only color singlets are allowed to propagate along links and all
Wilson loops for non-singlets average to zero.  We argue that, if this
model is to give QCD in its continuum limit, it must have a phase
transition.  We give arguments to support presence
of such a phase
transition.

\end{abstract}

\pacs{ }
\narrowtext
The large $N$ expansion is one of the few analytical tools available
for gauge
theories in the strong coupling regime.  However, in greater than two
dimensions
the solution of even the leading, $N=\infty$, order of the expansion has
not been found.  Recently, an intriguing new
approach to this problem has been proposed by Kazakov and Migdal
\cite{kazakmig}.  They have suggested using a lattice gauge model in
which the Yang-Mills interaction is induced by a heavy scalar field in
the adjoint representation of SU(N).  The action is
\equation
S = \sum_{x} N {\rm Tr} \bigl[
 m_{0}^{2} \Phi^{2}(x) - \sum_{\mu} \Phi(x)U_{\mu}(x)\Phi(x+\mu)
U^{\dagger}_{\mu}(x)\bigr)
\endequation
where the scalar $\Phi(x)$ lives on lattice sites and the link
operator $U_\mu(x)$ is the usual SU(N) group element-valued lattice
gauge field (in the fundamental representation).  Integrating over the
scalar field $\Phi$
\equation
\int DU D\Phi exp (-S) \sim \int DU exp ( -S_{ind}[U])
\endequation
results in the induced gauge action
\equation
S_{ind}[U] = - \frac{1}{2} \sum_{\Gamma} \frac{ |{\rm Tr} U[\Gamma]|^{2}}
{l[\Gamma] m_{0}^{2l[\Gamma]}}
\endequation
where $l[\Gamma]$ is the length of the loop $\Gamma$, $U[\Gamma]$ is
the ordered product of link operators along $\Gamma$ and the summation
is over all closed loops.  It was argued in \cite{kazakmig} that in the
continuum limit this action is equivalent to the ordinary Yang-Mills
action with coupling constant depending on the scalar mass and
ultraviolet cutoff.  Indeed, for an elementary plaquette, $\Box$, $
{\rm Tr}U[\Box]\sim N-{\rm Tr}F^2/2+\dots$ where $F$ is
the continuum field
strength and $S[U]\sim {\rm const}+{\rm Tr}F^2/4+\dots$ in the naive
continuum limit.  Furthermore, Kazakov and Migdal\cite{kazakmig}
showed that, when the gauge field is integrated first, one obtains an
effective matrix model for the field $\Phi(x)$ which can be analyzed
at large N and behaves like a master field in that its fluctuations
are frozen in the large N limit.  They conjectured that $\Phi(x)$ is
the correct master field for QCD.

In this Letter we shall point out that the lattice gauge theory
described by (1) has unusual properties which are not shared by the
usual Wilson formulation of lattice QCD.  This is a result of the fact
that the induced action (3) for the gauge fields depends only on the
modulus of the terms ${\rm Tr} U(\Gamma)$ and, unlike the Wilson theory, is
insensitive to their phases.  This is of course connected with the
fact that, since the scalar field transforms in the adjoint
representation of SU(N), the original action (1) has both a $U$ and a
$U^{\dagger}$ on each link.  The action has two local gauge
symmetries: the first one is usual local $SU(N)$ which acts on the
both the scalar and gauge field as left and right group multiplication
\equation
U_{\mu}(x) \rightarrow \Omega(x) U_{\mu}(x) \Omega^{-1}(x + \mu) ~~,~~
\Phi(x) \rightarrow \Omega(x) \Phi(x) \Omega^{-1}(x); \;\;
 \Omega(x)\in SU(N)
\endequation
and whose group elements are defined on the sites $x$.  The second
local symmetry is a hidden $Z_{N}$ symmetry which is not seen in the
weak coupling continuum limit and which is also absent in the standard
Wilson formulation of lattice QCD:
\equation
U_{\mu}(x) \rightarrow Z_{\mu}(x) U_{\mu}(x) \;\; Z_{\mu}(x) \in Z_{N}
\endequation
where $Z_N$ is the center of SU(N) and the local gauge group elements
are defined on links.  We stress that this symmetry will appear for
any induced $QCD$ if the original matter fields are invariant under
the action of the center of the gauge group (as is the adjoint
representation which we use here).

The properties of the strong coupling (confining) phase in a theory
with this additional local symmetry differ from those ordinary Wilson
lattice QCD - in some sense confinement is stronger and we call the
strong coupling phase {\it local confinement}.  Let us consider the
vacuum average of the Wilson line operator in the fundamental
representation
\equation W(C) =<{\rm Tr} P exp ( i \oint _{C} A_{\mu}
dx^{\mu})> = < {\rm Tr}
\prod_{\gamma \in C} U(\gamma)>
\endequation
 Using the local $Z_{N}$ symmetry $U(\gamma) \rightarrow e^{2\pi i
n_{\gamma}/N}U(\gamma)$ and assuming $Z_{N}$ invariance of the ground
state we get $W(C) = 0$ for any contour $C$ with non-zero area. A
non-zero result can only be obtained when in a Wilson loop or array of
Wilson loops, there is either an equal number of $U$ and $U^{\dagger}$
factors $\gamma$ ($\cdots U(\gamma) \cdots U^{\dagger}(\gamma)\cdots$)
or there are `baryon' factors $\prod_{1}^{N} U(\gamma)$ for every
link.  Thus we see that local $Z_{N}$ symmetry prohibits the
propagation of unscreened colour along the links - the only possible
type of excitations in this theory are `locally' white objects -
contrary to Wilson QCD where confinement was not so restrictive and
the Wilson line behaves as $\exp (- k {\cal A}(C))$ with string
tension $k$.  One can say that local $Z_{N}$ symmetry makes the string
tension infinite.

We also note that the $Z_N$ fluxons \cite{zn} whose contributions are
relevant to the Wilson theory are gauge artifacts in the present model
with local $Z_N$ symmetry.  Indeed, the flux through a plaquette is
defined modulo elements of the center of the group - due to the local
$Z_{N}$ symmetry one can change the product $\prod_{\gamma \in C}
U(\gamma)$ by the phase factor $exp (2\pi in/N) \in Z_{N}$.

The strong coupling phase of this theory exhibits local confinement.
An interesting question is whether there can be a phase transition to
a conventional confining phase or even a confinement-deconfinement
phase transition in this theory. In usual lattice gauge theory the
latter transition \cite{pol} is thought to occur at some value
of the gauge coupling when the number
of links $N_t$ in one of the Euclidean directions is finite.
It is connected with spontaneous breaking of a {\bf global} $Z_{N}$
symmetry and the appearance of a non-zero vacuum expectation value of
the relevant order parameter - the Polyakov line $<L> = < {\rm Tr}
\prod_{l=1}^{N_t}U(l)>$.  In the confining phase $<L> =
0$, in the deconfined phase there are $N$ degenerate states ( global
$Z_{N}$ is broken) and $<L> \in Z_{N}$.

However in our case besides the usual global $Z_{N}$ symmetry
we also have a {\bf local} $Z_{N}$ which will guarantee
that $<L> = 0$.  A local gauge symmetry cannot be spontaneously
broken\cite{Elit} and whether we can obtain a Wilson QCD-like phase of
this theory and a confinement-deconfinement transition within that
phase is a subtle question.  This question is an important one for the
Kazakov-Migdal model and for all induced QCD models where one starts
with the action without any explicitly symmetry breaking terms.

To better understand this issue, consider the following model of
lattice QCD with the gauge group SU(N) and with an action\cite{hamilton}
 \equation
S[U] = -\sum_{\Box}\left(\beta {\rm Re~ Tr} U(\Box) +
 \frac{1}{2\lambda} |{\rm Tr} U(\Box)|^{2} \right)
\endequation
where $\Box$ are fundamental plaquettes.  The first term is the
conventional Wilson action and the second term is the local $Z_N$
symmetric  term for elementary plaquettes which appears in (3).
The Wilson term breaks the local $Z_{N}$ symmetry explicitly. This
theory was considered in \cite{mak} in the limit $N \rightarrow
\infty$ where the possibility of phase transitions between the
confining phase and local confining phase (it was called the `absence
of quarks' phase) was discussed.

The $\beta\rightarrow0$ limit of (7) has local $Z_N$ symmetry and can
be regarded as a toy version of the Migdal-Kazakov model with the
identification $\lambda=4m_0^8$.  In the strong coupling (large
$\lambda$) limit it is a locally confining theory.  In its weak coupling
 limit it resembles continuum QCD in the sense that it produces the
correct naive continuum limit. The crucial question is whether there is,
in fact, a phase transition at some critical value of $\lambda_c$ of $\lambda$
to a Wilson (confining) phase or whether one must approach the continuum
limit from some nonzero value of $\beta$ in order to obtain conventional
QCD.  We shall present arguments that suggest that even for $\beta=0$
there is such a phase transition.
We speculate that for $\lambda<\lambda_c$
the $Z_N$ symmetry is realized in a Higgs phase and the physical properties
 of that phase resemble those of Wilson's lattice QCD.

To identify the Higgs field, we use a Gaussian
transformation to write the partition function as
\equation
Z = \int DU \exp\left(\sum_{\Box} \beta{\rm Re~Tr} U(\Box) +
\frac{1}{2\lambda} |{\rm Tr} U(\Box)|^{2} \right) =
\endequation
$$ = \int DU D\phi\exp \sum_{\Box}\left(- 2
\lambda|\phi(\Box)|^{2} + \phi(\Box){\rm Tr}U(\Box) +
\phi^{\dagger}(\Box){\rm Tr}U^{\dagger}(\Box) + 2\beta \lambda {\rm
Re}~\phi(\Box)\right)
$$
Here $\phi(\Box)$ is a scalar field which lives on plaquettes.  It is
a singlet under the gauge transformation in (4).  The local $Z_N$
transformations (5) act on
links. $\phi(\Box)$ transforms as
\equation
\phi(\Box)\rightarrow\phi(\Box)\prod_{{\rm links}\in\delta\Box}(Z_N)^{\pm1}
\endequation
where $\pm$ depends on the orientation of the link in the boundary
$\delta\Box$ of $\Box$.
The Wilson term in (7) results in a constant external field for
 $\phi(\Box)$ in (8) which breaks local $Z_N$ explicitly.

We wish to investigate the possibility that the local $Z_N$ symmetry
is realized in a Higgs phase. To this end we consider the effective
action for the Higgs field which is obtained by integrating the gauge fields in
(8),
$$
V^\beta_{\rm eff}=\sum_{\Box}\left(
2\lambda\vert\phi(\Box)\vert^2- 2\beta \lambda {\rm
Re}~\phi(\Box)\right)
$$
\equation
-\ln \int DU  \exp\left(
 \sum_{\Box}( \phi(\Box){\rm Tr}U(\Box)
 + \phi^{\dagger}(\Box){\rm Tr}U^{\dagger}(\Box)\right)
\endequation
Note that the integral is just the partition function of conventional QCD with
 a position dependent coupling constant given by $1/g^2\sim\phi(\Box)$.

For small $\phi$ we can use the conventional
 strong coupling expansion to evaluate $S_{\rm eff}$:
$$
S_{\rm eff}(\Box)= -
\ln \int DU  \exp\left(
 \sum_{\Box}2{\rm Re} \phi(\Box){\rm Tr}U(\Box)
 \right)
$$
$$
=\sum_{\Box,\Box'}\phi(\Box)\phi(\Box')
{\partial\over\partial\phi(\Box)}{\partial\over\partial\phi^*(\Box')}
S_{\rm eff}\vert_{\phi=0}+\dots
$$
\equation
=-\sum_{\Box\Box'}\left( <{\rm Tr}U(\Box) {\rm Tr}U^{\dagger}(\Box')>
- <{\rm Tr}U(\Box)><{\rm Tr}U(\Box')>\right)\phi(\Box)\phi(\Box')+\dots=
\endequation
$$
=-\sum_{\Box}\vert\phi(\Box)\vert^2+\ldots
$$
where the averaging $<..>$ is done at $\phi = 0$, i.e. it is simple
 integration $\int DU$ with zero action. To obtain the leading term
above note
 that $<{\rm Tr}U(\Box)> = 0$ and that  $<{\rm Tr}U(\Box) {\rm Tr}U^{\dagger}
(\Box')>=\delta_{\Box\Box'}$.
Higher order terms depend on either  higher powers of $\vert\phi\vert$
or on $\phi^N$ and on $\phi^{\dagger N}$.
Terms such as $\phi^N$ appear since the effective action is invariant
only under $Z_N$ but not under $U(1)$. ``Gradient'' terms appear
at order $\phi^6$ due to a term in the effective action which
is the product of $\phi$ over the six plaquettes on the face of any
cube. Such a product is invariant under local $Z_N$.
Note that higher order terms in $\vert\phi\vert$ will have positive signs
 and the potential appears not to be bounded from below.  Therefore, this
 expansion is good only for small $\vert\phi\vert$.
If we assume that
the expansion  has some non-zero radius of convergence  the effective potential
 for $\phi$ in the small $\phi$ approximation is (for $\beta=0$)
\equation
V_{\rm eff}[\phi] = 2\lambda |\phi|^{2} - |\phi|^{2} + \ldots
\endequation
which exhibits the typical behaviour of a phase transition to a Higgs phase at
$\lambda
 = 1/2$. This suggests that
for large $\lambda$ this model exists in a locally confining phase
and that at some critical value $\lambda_c$ of $\lambda$
there is a phase transition to what we conjecture is
a conventional confining phase resembling Wilson lattice QCD.

There is another way to demonstrate that
for small $\lambda$ one can get the non-zero vacuum expectation value for the
master
 field $\phi(\Box)$. Asymptotically, when $\phi$ is large,
 the logarithm of the integral in (10) is bounded using the triangle
inequality,\equation
\int dU\exp\sum_\Box(2{\rm Re}\phi{\rm Tr}U)\leq\exp(\sum_\Box 2\vert\phi\vert)
\int dU=\exp(\sum_\Box 2\vert\phi\vert)
\endequation
so that for large $\vert\phi\vert$,
\equation
S_{\rm eff}=\sum_{\Box}\left(
2\lambda\vert\phi(\Box)\vert^2- 2\vert\phi(\Box)\vert-2\beta \lambda {\rm
Re}~\phi(\Box)\right)
\endequation
The effective potential is extremized by the configuration
$\phi=\beta/2+{\rm sign}(\phi)/2\lambda$.  When $\beta<1/\lambda$ there
are two solutions which are degenerate at  $\beta=0$.  This non-trivial
solution is reliable for small $\lambda$ since the estimate of the effective
potential is valid for large $\phi$.
 The nature of the solution changes when the external field $\beta>1/\lambda$.
   Note that our results are in qualitative agreement
 with the analysis \cite{mak}.

It is unclear how to generalize these arguments  to the case
 of the  induced action (3)
 $S_{ind}(U)$ where we have sum of all possible closed paths,
 not only the minimal path  as in our toy model.
One option is to  consider the sum
 of different master fields $\phi_{\Gamma}$ (analogous to a string field)
and if even one of them
  has a non-zero expectation value, the local $Z_{N}$ symmetry may
 be realized in the in the Higgs mode.
It is easy to see  that for each master field
$\phi_{\Gamma}$ the effective coupling
$\lambda_{\Gamma} =
l[\Gamma] m_{0}^{2l[\Gamma]}$ and the leading quadratic term in the effective
 action is like $\sum_{\Gamma}(2\lambda(\Gamma)-1)\vert\phi(\Gamma)\vert^2$.
If
 one decreases $m_{0}$  it will be the the coupling constant corresponding to
the
 smallest loop, i.e. $\lambda_{\Box}$,  which becomes smaller than $1/2$ first.
 One can imagine
that if all of the other coupling constants are larger than the corresponding
 critical values the effective theory will be the Wilson theory with
 the fundamental plaquette action.

We believe that one of the principal questions about induced QCD
with local $Z_{N}$
 symmetry is whether there is a weak coupling phase with ordinary confinement
 and an area law for Wilson loops.  In this paper we have
 speculated about such a possibility  based on the assumption that the
 Higgs phase for our auxiliary model is equivalent to a confinement phase.
 To better understand how in the Higgs phase for $\phi$ one can get an
 area law  for Wilson loops
let us consider the generalization of the usual Wilson loop
\equation
W_{\phi}(C) = < W(C) \prod_{\Box \in A(C)} \phi(\Box)>
\endequation
We have introduced this
 ``filled'' Wilson line in which we preserve the local $Z_{N}$
 symmetry by  introducing the product of factors $\phi(\Box)$
 for each plaquette in the interior of the contour $C$. Now one can see that
if the master field $\phi$
 acquires a non-zero vacuum expectation value $<\phi>$ one gets
\equation W_{\phi}(C) \sim
exp(A(C) \ln <\phi>)<W(C)>_{\phi=<\phi>}
\endequation

It is interesting to note that these two phases may be
 equivalent (or at least related) to the two phases of the matrix model
obtained
 by Kazakov and Migdal \cite{kazakmig}. They found
 that in the weak coupling case, i.e. when $m_{0} <<1$
 induced QCD can be reduced to the two-matrix
 model
\equation
I(\Phi~~\Psi) \sim
 \int d^{N}xd^{N}y \Delta(x)\Delta(y)
 exp\bigl(N\sum_{i} x_{i}y_{i} - V_{\Phi}(x_{i}) - V_{\Psi}(y_{i})\bigr)
\endequation
with potentials
\equation
V_{\Phi,\Psi}(x) = \int d\mu \rho_{\Phi,\Psi}(\mu) \ln (\mu - x)
\endequation
and a gap in the  eigenvalue distribution function
 $\rho(\mu)$.
In the strong coupling phase , i.e. when $m_0$ is large and in the framework of
a $1/m_{0}$
 expansion, there is no gap, the potential becomes singular and
 the system is in a new phase.  This strong coupling phase was considered by
Migdal
 in a recent preprint \cite{migdal}, where he obtained a critical
 $m_{0}$ with numerical value $m^{2}_{c}  = 4$ for a four-dimensional
 lattice ($D$ in general case , but only at $D=4$ does the induced action
 have a naive local Yang-Mills limit).

 We conjecture that the phase transition
 in $m_{0}$ in the
 Kazakov-Migdal matrix model   is precisely the phase transition between
 the local confinement and  confinement phases and that the strong coupling
phase
 corresponds to the local confinement with unbroken $Z_{N}$ and
 weak coupling phase corresponds to the restoration of area law and
(some kind of) spontaneous breaking of $Z_{N}$ symmetry (Higgs
 phase for $\phi$ master field). It is very natural to
 conjecture also that the gap in the eigenvalues distribution
 function  $\rho(\mu)$ is proportional to some inverse power of
 the string tension $\alpha'$. Then in a weak coupling phase
 the gap is non-zero as well as $\alpha'$ and one recovers the
 area law. However in the strong coupling case the gap disappears
 and string tension diverges $\alpha'\rightarrow \infty$, which
 means that we are in a local confinement  phase.  In a future publication
 we shall present the results of more detailed investigations.

\vfill
\eject

\nonum\section{acknowledgments}
This work was supported in part by the Natural Sciences and Engineering Council
of
Canada. We thank A. Morozov for discussions.


\vspace{13cm}

\begin{references}
\bibitem{hamilton} It is also possible to consider a Hamiltonian
version of QCD with the same local $Z_N$ symmetry. If we write
$H=\sum g^2E^2-\sum{(1/g^2)}\vert{\rm Tr}U(\Box)\vert^2$ then $H$
has the same naive continuum limit as ordinary Hamiltonian lattice
QCD but it also has the additional local $Z_N$ symmetry.
This model differs in some ways from the Euclidean theory.
In the Hamiltonian version one can satisfy Gauss' Law by joining
a quark and antiquark source with a string. In strong coupling
this will lead to and energy which grows linearly with separation.
This is in contrast with the Euclidean version for which the
Wilson loop vanishes identically.
\end{references}
\end{document}